\documentclass[a4paper,11pt]{article}
\usepackage{pos}

\title{Medium-scale anisotropies measured by Telescope Array surface detectors}
\ShortTitle{TA SD Anisotropies}

\author*[a]{Jihyun Kim}
\author[a]{Dmitri Ivanov}
\author[b]{Kazumasa Kawata}
\author[b]{Hiroyuki Sagawa}
\author[a]{Gordon Thomson}

\affiliation[a]{High Energy Astrophysics Institute and Department of Physics and Astronomy, University of Utah\\
Salt Lake City, Utah 84112-0830, USA}
\affiliation[b]{Institute for Cosmic Ray Research, University of Tokyo,\\
Kashiwa, Chiba 277-8582, Japan}

\onbehalf{on behalf of the Telescope Array Collaboration} 

\emailAdd{jihyun@cosmic.utah.edu}

\abstract{The Telescope Array (TA) experiment, the largest observatory for ultra-high energy cosmic rays in the Northern Hemisphere, has identified two medium-scale anisotropies: the TA Hotspot near the constellation Ursa Major and an excess in the direction of the Perseus-Pisces supercluster. Studying these medium-scale anisotropies may provide insights into the origins of ultra-high energy cosmic rays. This presentation will explore an oversampling analysis of TA surface detector data to evaluate these medium-scale event excesses and will present the latest findings on the TA Hotspot and the Perseus-Pisces supercluster excess.}

\ConferenceLogo{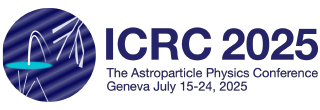}

\FullConference{39th International Cosmic Ray Conference (ICRC2025)\\
 15–24 July 2025\\
Geneva, Switzerland\\}
\begin{document}
\maketitle

\vspace{-3mm}
\section{Introduction}
\label{sec:intro}
\vspace{-3mm}
Charged particles of extremely high energy, known as ultra-high energy cosmic rays (UHECRs), arrive from outer space carrying energies that exceed $10^{18}$~eV. Upon striking the Earth's atmosphere, these primary particles interact with atmospheric atomic nuclei, which initiates an extensive cascade of secondary particle production. This cascading process ultimately results in atmospheric particle showers comprising millions of individual subatomic components.

The Telescope Array (TA) is designed to detect these extensive air showers initiated by UHECR particles. Situated in Utah’s western desert in the USA, at geographic coordinates $39.3^\circ \mathrm{N}$, $112.9^\circ \mathrm{W}$, it stands as the largest UHECR observatory in the Northern Hemisphere. Located at an elevation of 1,400 meters above sea level, the site is ideal for capturing the peak development of extensive air showers. 

TA employs a hybrid detection technique, combining a surface detector (SD) array with fluorescence telescopes. The SD array consists of 507 individual counters, each spaced 1.2 km apart, collectively covering an area of approximately 700 km$^2$. Each detector features two layers of plastic scintillators designed to register the particle footprint of air showers at ground level~\cite{TelescopeArray:2012uws}. Complementing this, three fluorescence detector (FD) stations, equipped with a total of 38 fluorescence telescopes, monitor the sky above the SD array across an elevation angle of $3^\circ$--$31^\circ$. These telescopes are sensitive to the ultraviolet emissions produced as UHECR-induced air showers propagate through the atmosphere~\cite{Tokuno:2012mi}.

We analyze the properties of extensive air shower events measured by the SD and FD systems to reconstruct the fundamental characteristics of the original particles, such as their energies, mass compositions, and arrival directions. By investigating the nature and origins of UHECRs, the TA aims to identify their sources and understand the mechanisms responsible for their extreme energies, thereby offering insights into some of the most energetic and violent phenomena in the universe. 

To identify the astrophysical sources of UHECRs, we search for non-uniform patterns in the spatial distribution of cosmic ray arrival directions, looking for evidence of directional clustering rather than a random, isotropic distribution. Such anisotropic patterns could reveal the locations of sources and the acceleration mechanisms behind these particles. This investigation is exceptionally challenging, as cosmic magnetic fields systematically deflect particle trajectories. Nevertheless, ongoing research continues to advance our understanding of the plausible astrophysical sites that may host UHECR sources.

In 2014, the TA collaboration published indications of medium-scale anisotropy in the arrival directions of UHECRs with energies greater than $5.7\times10^{19}$~eV, a feature that became known as the TA Hotspot~\cite{TelescopeArray:2014tsd}. Analysis of five years of data from the TA SD array revealed a concentration of events near the Ursa Major constellation. The investigation employed an oversampling method with a $20^\circ$ angular distance window to identify this clustering pattern. Statistical evaluation using the Li-Ma method~\cite{Li:1983fv} quantified the maximum local significance as $5.1\sigma$ at the sky position $(146.7^\circ, 43.2^\circ)$ in equatorial coordinates. Through extensive Monte Carlo simulations, the post-trial significance was estimated to be $3.4\sigma$.

Several potential astrophysical sources have been proposed to explain the hotspot, including the galaxies M82 and Mrk 180~\cite{He:2014mqa}, the active galactic nucleus Mrk 421~\cite{Fang:2014uja}, and galaxy filaments connected to the Virgo Cluster~\cite{Kim:2019eib}. However, none of these candidates has been definitively confirmed, which indicates that further observational data will be necessary to resolve the nature and origin of the TA Hotspot.

In more recent work~\cite{TelescopeArray:2021dfb, Kim:2023xfc, Kim:2023ksw}, the TA \textcolor{black}{collaboration} has identified another statistically significant excess among events with energies exceeding $10^{19.4}$~eV. This finding emerged during investigations into differences in energy spectrum measurements between TA and the Pierre Auger Observatory (Auger). When examining the arrival directions of events in this relatively lower energy range, an excess of events was found aligned with the Perseus-Pisces Supercluster (PPSC) structure. This feature is now referred to as the PPSC excess.

In this study, we report on the investigation of medium-scale anisotropies, specifically, the TA Hotspot and the PPSC excess, using the 16-year dataset collected by the Telescope Array surface detector array between May 11, 2008, and May 10, 2024.

\vspace{-3mm}
\section{Oversampling Analysis Methods}
\label{sec:analysis}
\vspace{-3mm}
To investigate medium-scale anisotropies in the data, we perform oversampling analyses across a grid in equatorial coordinates. At each grid point, we count the number of events within a predefined angular distance window, denoted as $N_{\rm on}$. The complementary count, $N_{\rm off} = N_{\rm tot} - N_{\rm on}$, is derived from the total number of events, $N_{\rm tot}$, in the data set.

We then repeat this procedure using $10^5$ simulated cosmic ray events generated under the assumption of an isotropic flux. These simulations incorporate the geometrical exposure function $g(\theta) = \sin \theta \cos \theta$, where $\theta$ is the zenith angle, assuming a uniform detection efficiency across all zenith angles in the relevant energy range. From the isotropic sample, we define the exposure ratio $\alpha = N_{\rm iso, on}/N_{\rm iso, off}$, where $N_{\rm iso, on}$ and $N_{\rm iso, off}$ represent the number of simulated events inside and outside the angular distance window, respectively. By comparing the observed oversampling results to the isotropic expectation, we compute the statistical significance of any excess using the Li-Ma formula~\cite{Li:1983fv}:
\begin{equation}
    S_{\rm LM}=\sqrt{2} \, \left[ N_{\rm on} \ln{ \left( \frac{(1+\alpha) N_{\rm on}} 
    {\alpha(N_{\rm on}+N_{\rm off})} \right) }
    + N_{\rm off} \ln{ \left( \frac{(1+\alpha) N_{\rm off}}{N_{\rm on}+N_{\rm off}} \right)} \right]^{1/2}.
\end{equation}

\vspace{-3mm}
\section{Results}
\label{sec:anisotropy}
\vspace{-3mm}
\begin{figure}[t]
    \centering
    \includegraphics[width=0.6\columnwidth]{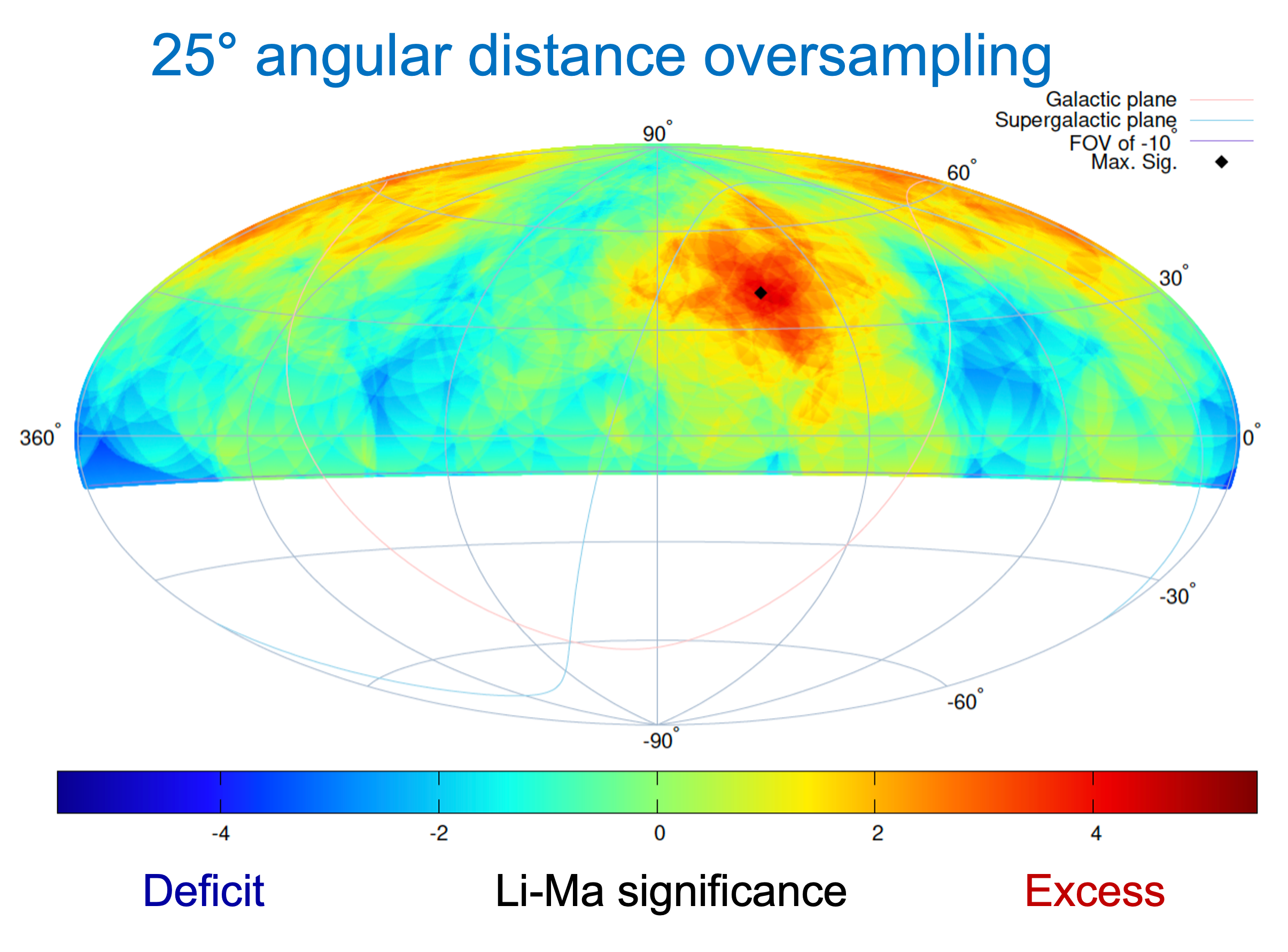}
    \caption{\textbf{The sky map of the TA Hotspot using Hammer projection.} The Li-Ma significance using a $25^\circ$-circle angular window is shown in equatorial coordinates for 16 years of SD events with energies greater than $5.7\times10^{19}$~eV. The black diamond indicates the maximum Li-Ma significance position measured at ($144.0^\circ,\, 40.5^\circ$). The color code indicates an excess (red) and a deficit (blue) of events compared to isotropy.}
    \label{hotspot_skymap} 
\end{figure}

\begin{figure}[t]
    \centering
    \includegraphics[width=0.95\columnwidth]{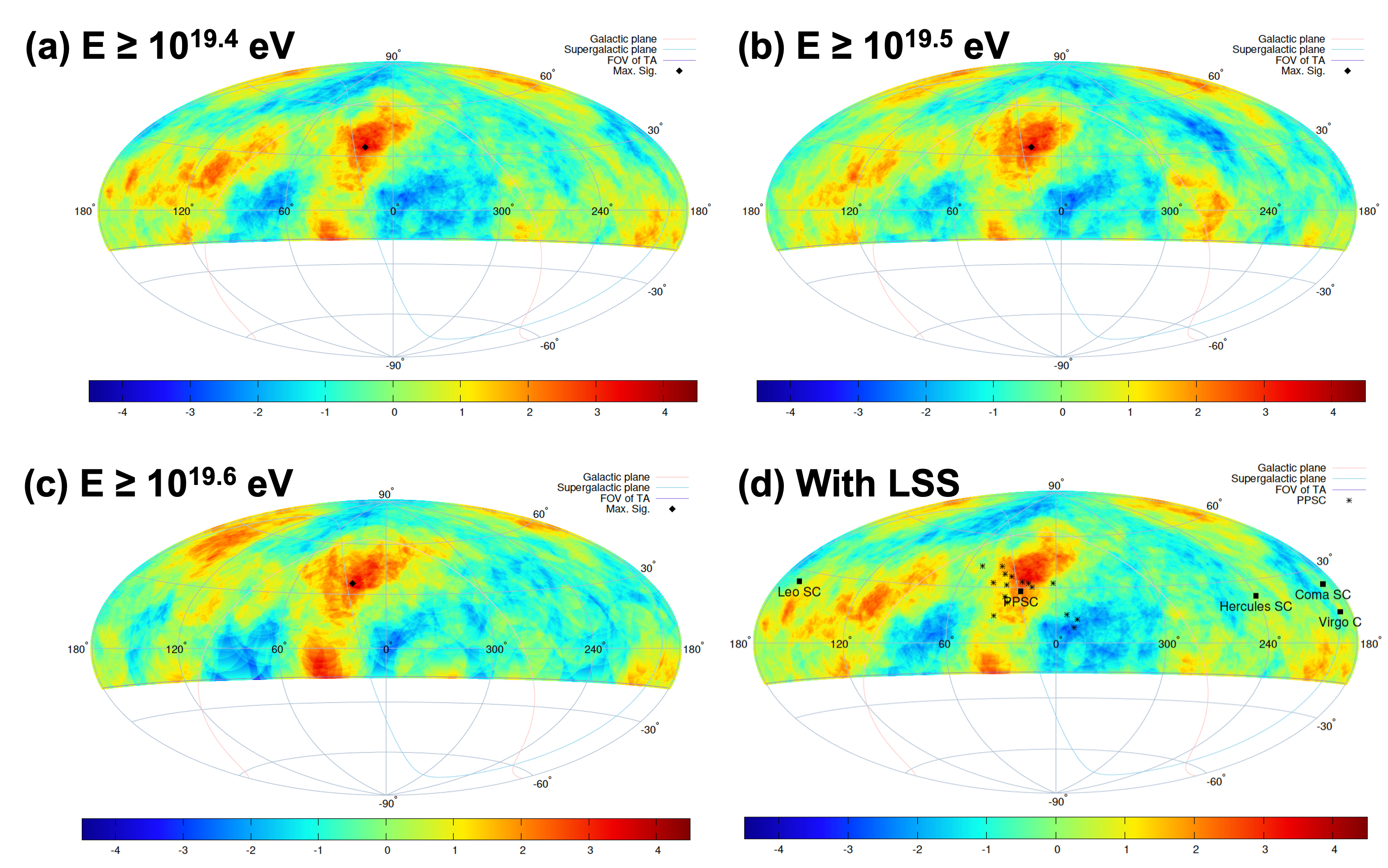}
    \caption{\textbf{The sky maps of the Perseus-Pisces Supercluster (PPSC) excess using Hammer projection.} The Li-Ma significance using a $20^{\circ}$-circle angular window is shown in equatorial coordinates for different energy thresholds: (a) $E \geq 10^{19.4}$~eV, (b) $E \geq 10^{19.5}$~eV, and (c) $E \geq 10^{19.6}$~eV. The black diamonds indicate the maximum Li-Ma significance positions for each energy threshold. Additionally, (d) shows the nearby major large-scale structures (LSS) overlaid with the Li-Ma significance map for $E \geq 10^{19.4}$~eV. The color code indicates an excess (red) and a deficit (blue) of events compared to isotropy.}
    \label{ppsc_skymap}
\end{figure}

\begin{figure}[ht]
    \centering
    \includegraphics[width=0.4\columnwidth]{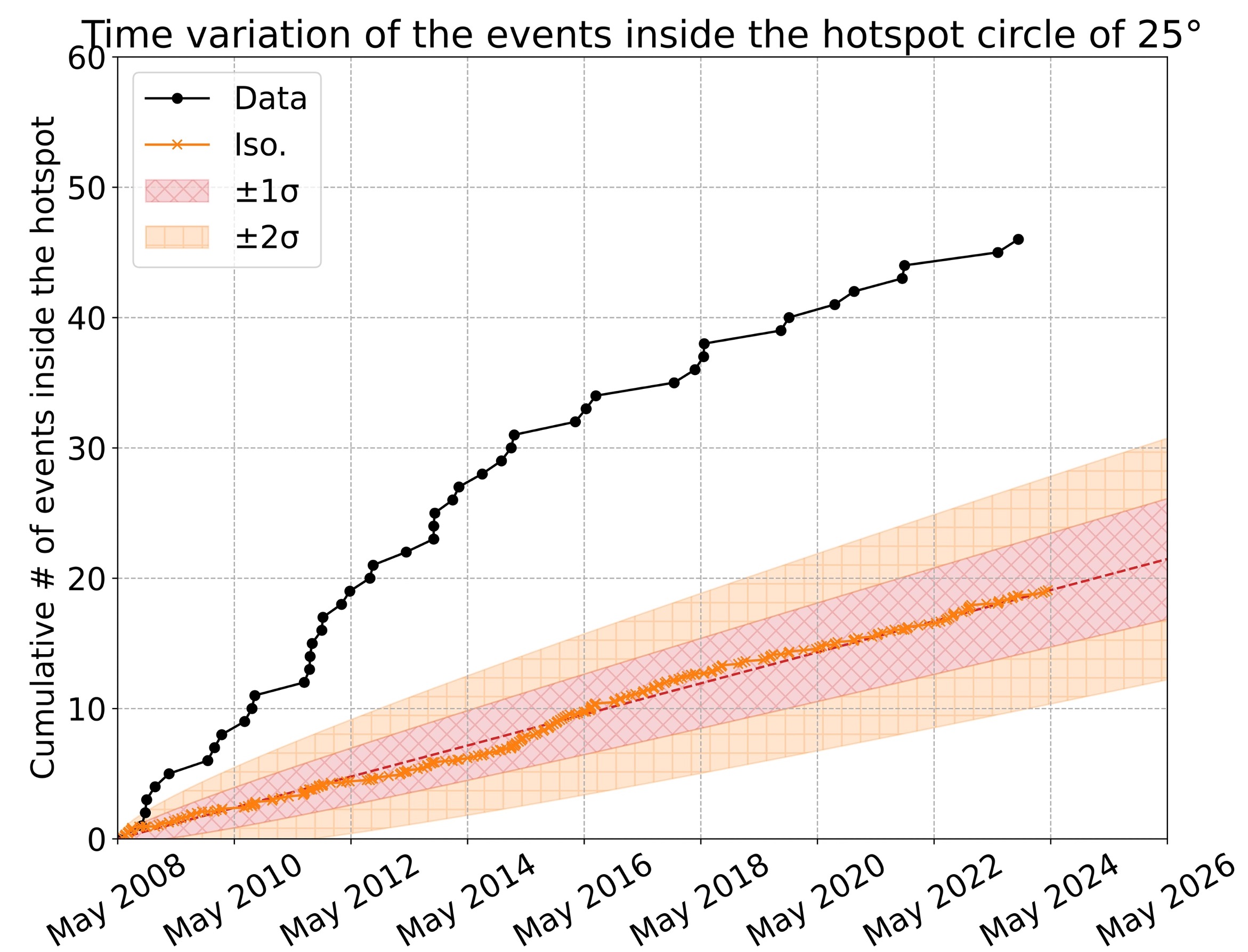}
    \includegraphics[width=0.4\columnwidth]{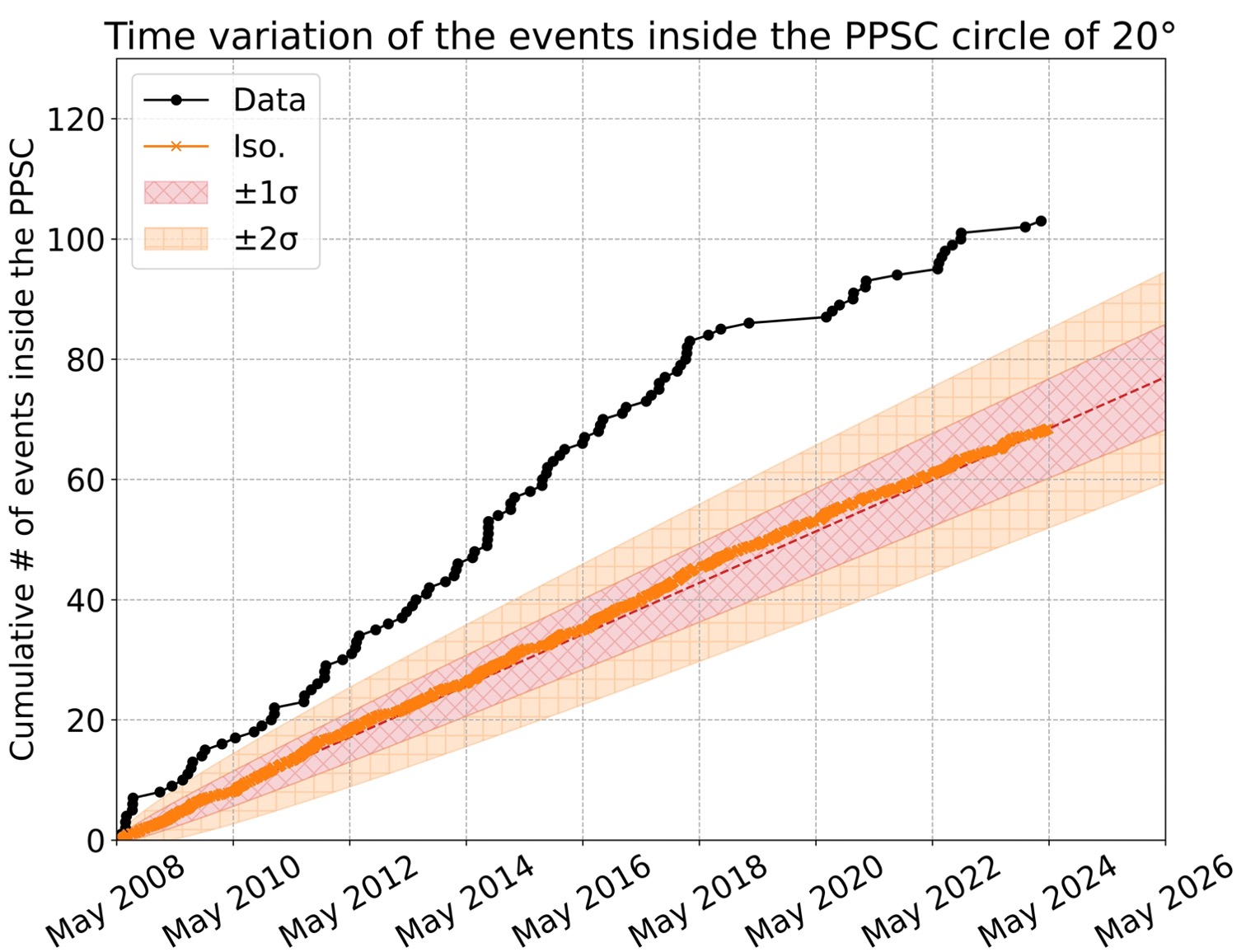}
    \caption{\textbf{Evolution of medium-scale anisotropies.} \textbf{(Left)} Cumulative number of events with $E>5.7\times10^{19}$~eV (black dots) within a $25^\circ$ radius centered at the Hotspot's maximum Li-Ma significance ($144.0^\circ,\, 40.5^\circ$), compared to the isotropic expectation (orange X's). The red dashed line represents the estimated isotropic event rate, with shaded bands indicating $\pm 1\sigma$ (pink) and $\pm 2\sigma$ (orange) statistical fluctuations. \textbf{(Right)} Cumulative number of events with $E>10^{19.4}$~eV (black dots) within a $20^\circ$ radius around the PPSC excess center ($17.9^\circ$, $35.2^\circ$), shown alongside the isotropic expectation (orange X's) and the same statistical bands.}
    \label{HotspotPPSCGrowth}
\end{figure}

\subsection{TA Hotspot}
The following event selection criteria were applied in the TA Hotspot analysis~\cite{TelescopeArray:2014tsd}: (1) each event had to include at least four SD counters, (2) the zenith angle of the event's arrival direction had to be less than $55^{\circ}$, and (3) the reconstructed energy had to exceed $5.7\times10^{19}$~eV. This energy threshold was adopted based on the correlation study with active galactic nuclei conducted by the Auger \textcolor{black}{collaboration}~\cite{PierreAuger:2007pcg}, in order to avoid introducing a free parameter during the scan of the phase space.

Oversampling analyses were performed at grid points defined in equatorial coordinates, with a step size of $0.1^\circ$ in both right ascension (ranging from $0^\circ$ to $360^\circ$) and declination (from $-10^\circ$ to $90^\circ$). An angular window with a $20^\circ$ radius was initially used, following medium-scale anisotropy studies by the AGASA collaboration~\cite{AGASA:1998frs, AGASA:1999jxd}, to avoid arbitrary parameter scanning. Later, we included scans over \textcolor{black}{five different} window sizes and identified a $25^\circ$ radius as yielding the most significant excess. In recent analyses, this $25^\circ$ window has been adopted as a fixed parameter for consistency.

Here, we analyze 16 years of data collected by the TA SD array. This dataset comprises 228 events with energies exceeding $5.7\times10^{19}$~eV. Using a $25^\circ$ angular distance window, we find the most significant excess of events at equatorial coordinates ($144.0^\circ$, $40.5^\circ$), with a Li-Ma significance of $4.9\sigma$. Within this $25^\circ$ radius, 46 events are observed, whereas an isotropic distribution would predict 19.1 events. Figure~\ref{hotspot_skymap} shows the sky map of the TA Hotspot in equatorial coordinates displayed using a Hammer projection. The color scale represents deviations from isotropy, with red indicating excesses and blue indicating deficits compared to isotropic expectation. The black diamond marks the location of the highest Li-Ma significance at ($144.0^\circ$, $40.5^\circ$).

To assess the likelihood of such an excess arising by chance in an isotropic UHECR sky, we generate Monte Carlo simulations. Each simulated dataset contains the same number of events as observed, distributed isotropically while accounting for the geometrical exposure of the TA SD array. A trial is considered a success if the maximum Li-Ma significance in the simulation satisfies $S_{\rm MC} \geq S_{\rm obs}$. The resulting chance probability of observing such an excess anywhere in TA's field of view is found to be $2.1 \times 10^{-3}$, which corresponds to a significance of approximately $2.9\sigma$.

\subsection{Perseus-Pisces Supercluster Excess}

An additional excess was identified in events with slightly lower energies, specifically those above $10^{19.4}$~eV, during an investigation into the difference between the TA and Auger energy spectra in the common sky region~\cite{TelescopeArray:2018ygn, TelescopeArray:2024tbi}. Part of this study aimed to explore whether the TA Hotspot extends toward lower declinations at energies below the original Hotspot threshold. To maintain consistency with the original Hotspot analysis~\cite{TelescopeArray:2014tsd}, we employed a fixed angular window of $20^\circ$ in the oversampling procedure, thereby avoiding the introduction of a free parameter in the scan~\cite{TelescopeArray:2021dfb}. This approach led to the identification of an additional excess in the arrival direction distribution, known as the Perseus-Pisces Supercluster (PPSC) excess.

The event selection criteria used in the PPSC excess analysis are identical to those applied in the TA–Auger energy spectrum comparison study. This consistency allows for a unified investigation of both spectral features and anisotropy at lower energies, while preserving reliable energy and angular resolution: (1) each event must involve at least five SD counters, (2) the reconstructed primary zenith angle must be less than $55^\circ$, (3) both the geometry and lateral distribution fits must have a $\chi^2$/degree of freedom less than 4, (4) the angular uncertainty estimated by the geometry fit must be less than $5^\circ$, (5) the fractional uncertainty in $S(800)$ estimated by the lateral distribution fit must be less than 25\%, and (6) the counter with the largest signal must be surrounded by four working counters—one to the north, east, south, and west on the grid, although these counters do not need to be immediate neighbors of the largest signal counter.

Over 16 years of data collection with the TA SD array, we recorded 1186, 767, and 464 events with energies exceeding $10^{19.4}$~eV, $10^{19.5}$~eV, and $10^{19.6}$~eV, respectively. Oversampling analyses were performed at grid points in equatorial coordinates, with a step size of $0.1^\circ$ in both right ascension (ranging from $0^\circ$ to $360^\circ$) and declination (from $-15.7^\circ$ to $90^\circ$). The resulting Li-Ma significances for each energy threshold are as follows: $3.7\sigma$ at ($17.9^\circ$, $35.2^\circ$) for $E \geq 10^{19.4}$~eV, $3.9\sigma$ at ($19.2^\circ$, $35.2^\circ$) for $E \geq 10^{19.5}$~eV, and $3.7\sigma$ at ($21.8^\circ$, $36.2^\circ$) for $E \geq 10^{19.6}$~eV. These consistent excesses are observed in the direction of the PPSC, with angular separations of $7.7^\circ$, $7.4^\circ$, and $8.3^\circ$ from \textcolor{black}{the PPSC} center, respectively. The corresponding sky maps for the PPSC excess at each energy threshold are shown in Figure~\ref{ppsc_skymap}.

Figure~\ref{ppsc_skymap}(d) shows nearby large-scale structures similar to the PPSC, within the TA’s field of view and up to a distance of 150 Mpc, the GZK horizon for proton primaries, overlaid on the excess map. These structures include the Virgo Cluster (17 Mpc), PPSC (70 Mpc), Coma Supercluster (90 Mpc), Leo Supercluster (135 Mpc), and Hercules Supercluster (135 Mpc). No comparable excess is observed near any of these other major structures. The PPSC stands out as a unique and significant structure within the TA's field of view. It is the closest supercluster to Earth aside from the Local Supercluster and is located near the Local Void~\cite{Courtois:2013yfa, Tully:2019ngb}, where the magnetic field strength is expected to be weaker than in other regions of the cosmic web. The PPSC’s proximity, location near the Local Void, and the absence of similar excesses elsewhere make this observation particularly compelling. This further highlights the PPSC as a promising target for future studies aimed at identifying the sources of UHECRs. 

We also examine the evolution of the TA Hotspot and the PPSC excess. Figure~\ref{HotspotPPSCGrowth} displays the \textcolor{black}{time evolution of the} cumulative event counts for the TA Hotspot (left) and the PPSC excess (right). In each panel, black dots represent the total number of observed events within a $25^\circ$ radius centered at ($144.0^\circ$, $40.5^\circ$) for the Hotspot and a $20^\circ$ radius centered at ($17.9^\circ$, $35.2^\circ$) for the PPSC excess. The orange X markers denote the cumulative number of events expected under an isotropic distribution. The red dashed line indicates the expected isotropic event rate, while the shaded pink and orange bands correspond to $\pm 1\sigma$ and $\pm 2\sigma$ statistical fluctuations, respectively. This figure illustrates the accumulation of observed events deviating from the isotropic expectation and shows the temporal evolution of the statistical significance of these deviations.

\vspace{-3mm}
\section{TA Hotspot Visibility Under Exposure Weighting}
\label{sec:dec_weight}
\vspace{-3mm}

\begin{figure}[ht]
    \centering
    \includegraphics[width=0.8\columnwidth]{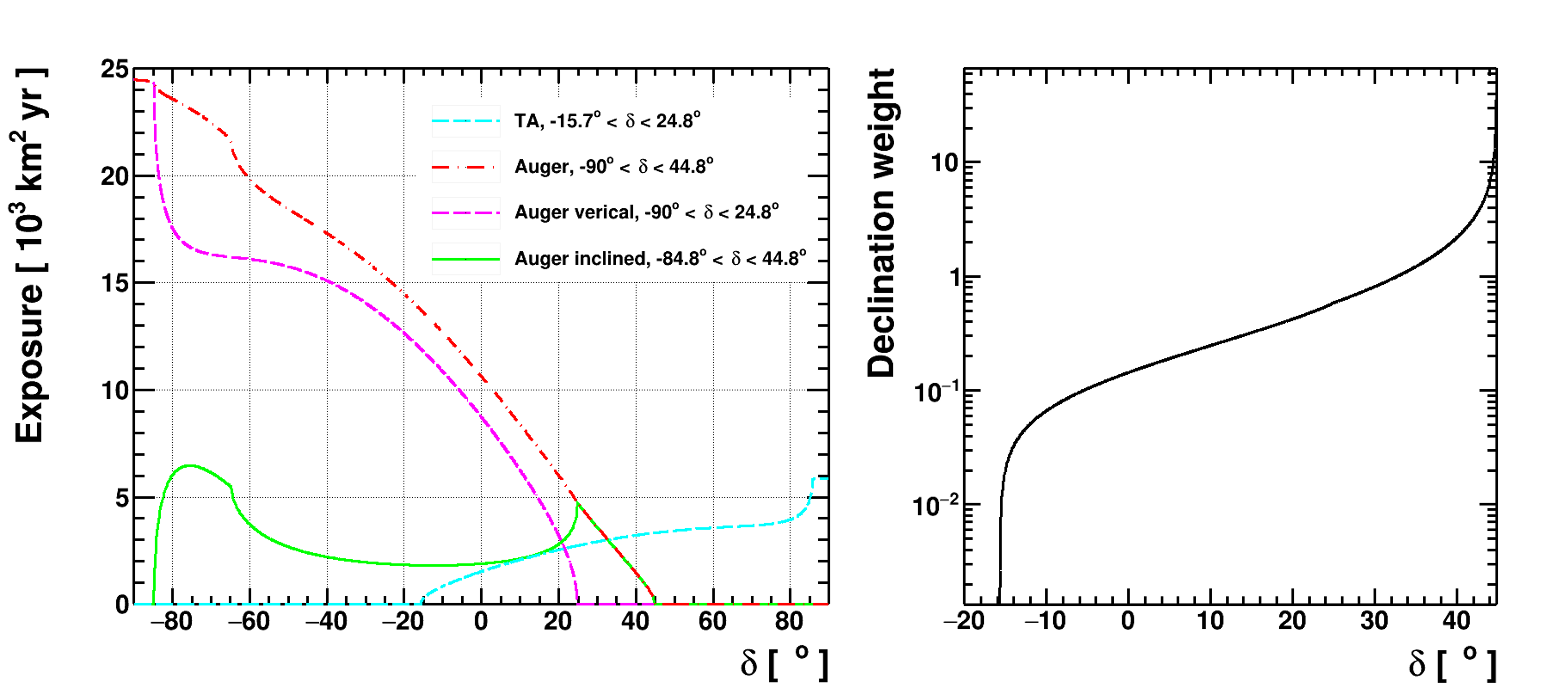}
    \caption{\textbf{TA and Auger exposures and exposure ratio of TA to Auger as a function of declination} \textbf{(Left)} Exposure distributions for the Auger vertical, Auger inclined, and TA spectra, highlighting the complementary coverage of the declination band. \textbf{(Right)} Exposure ratio of TA to Auger as a function of declination, illustrating the mismatch in coverage.}
    \label{fig:exposure_comparison}
\end{figure}

\begin{figure}[ht]
    \centering
    \includegraphics[width=1\columnwidth]{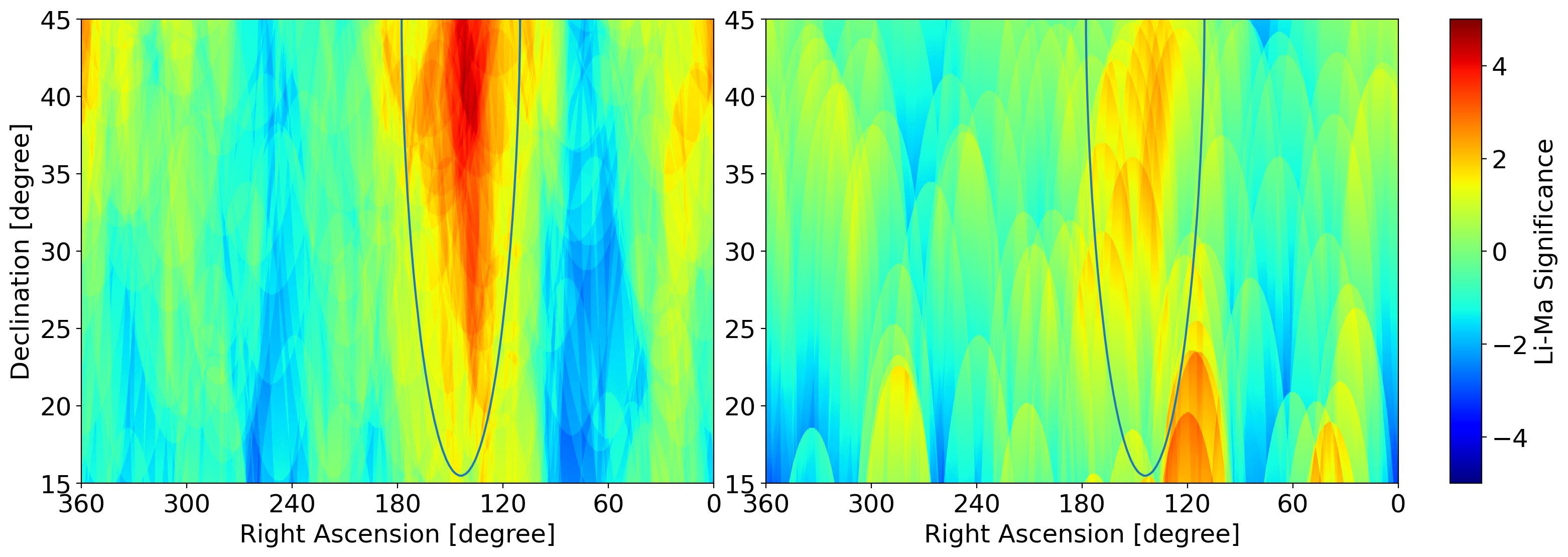}
    \caption{\textbf{Comparison of the TA Hotspot using declination weights in equatorial coordinates.} \textbf{(Left)} The TA Hotspot as observed by TA. \textbf{(Right)} The TA Hotspot as it would appear \textcolor{black}{in the Auger detector}, accounting for the declination weights of TA Hotspot events. The blue solid lines represent a $25^\circ$ angular distance window from the Hotspot center.}
    \label{fig:weighted_skymap}
\end{figure}

In this section, we discuss how significant the TA Hotspot would appear to Auger. In~\cite{PierreAuger:2024hrj}, the Auger Collaboration investigated the excess regions reported by TA using both vertical and inclined air showers, which now cover declinations up to $44.8^\circ$. They reported no excesses in the Auger data corresponding to the TA Hotspot and the PPSC excess regions.

Figure~\ref{fig:exposure_comparison} shows the TA and Auger exposures as a function of declination (left), and the ratio of TA to Auger exposures as a function of declination (right). 
%The ratio is heavily weighted at both edges of the declination band. 
While the total integrated exposures accumulated by Auger in these regions are comparable to those of TA, a direct comparison is not straightforward due to differences in the exposure profiles. Specifically, the Auger's exposure drops off rapidly toward the northernmost declinations, which significantly affects the sensitivity to anisotropic features such as the TA Hotspot. 

To evaluate how the TA Hotspot would appear to Auger, we weight the TA events according to the ratio of Auger to TA exposures as a function of declination. Figure~\ref{fig:weighted_skymap} shows the TA Hotspot \textcolor{black}{(as viewed by TA)} for reference (left), and the weighted TA events based on the exposure ratio (right). Here, we focus on the Hotspot region defined by $15^\circ < \delta < 45^\circ$. At the center of the TA Hotspot ($144.0^\circ, 40.5^\circ$), while TA observes a local Li-Ma significance of approximately $5\sigma$, the weighted significance drops to around $2\sigma$ as seen by Auger, i.e., statistically insignificant. Given the rapid decrease in exposure at high declinations, Auger is inherently less sensitive to anisotropies in the northern sky, making it unlikely to observe the excesses reported by TA with present exposure.

\vspace{-3mm}
\section{Summary}
\label{sec:sum}
\vspace{-3mm}

Using 16 years of data from the Telescope Array surface detector array, we examine medium-scale anisotropies in the arrival directions of UHECRs. Our analysis confirms the presence of a persistent Hotspot at the highest energies ($E \geq 5.7 \times 10^{19}$~eV) near the Ursa Major constellation, with a local significance of 4.9$\sigma$ and a global significance of 2.9$\sigma$. We also report an updated observation of an excess of events at slightly lower energies ($E \geq 10^{19.4}$~eV) in the direction of the Perseus-Pisces supercluster, exhibiting a local significance of 3.7$\sigma$ and a chance probability corresponding to 3.1$\sigma$ for such an excess occurring near the Perseus-Pisces supercluster.

These results are consistent with earlier measurements of medium-scale anisotropies and support a possible correlation between UHECR arrival directions and large-scale structures of the universe. This correlation may provide insights into the distribution of UHECR sources or the influence of intergalactic magnetic fields. The ongoing expansion of the Telescope Array experiment, TA$\times$4, is expected to substantially increase data collection capabilities, which play a crucial role in identifying the origins of UHECRs. Continued observations will be essential for further investigating the nature and stability of both the TA Hotspot and the PPSC excess.

% \bibliographystyle{JHEP}
% \bibliography{references}

\providecommand{\href}[2]{#2}\begingroup\raggedright\begin{thebibliography}{10}

\bibitem{TelescopeArray:2012uws}
{\scshape Telescope Array} collaboration, \emph{{The surface detector array of the Telescope Array experiment}}, \href{https://doi.org/10.1016/j.nima.2012.05.079}{\emph{Nucl. Instrum. Meth. A} {\bfseries 689} (2013) 87} [\href{https://arxiv.org/abs/1201.4964}{{\ttfamily arXiv:1201.4964[astro-ph.IM]}}].

\bibitem{Tokuno:2012mi}
H.~Tokuno et~al., \emph{{New air fluorescence detectors employed in the Telescope Array experiment}}, \href{https://doi.org/10.1016/j.nima.2012.02.044}{\emph{Nucl. Instrum. Meth. A} {\bfseries 676} (2012) 54} [\href{https://arxiv.org/abs/1201.0002}{{\ttfamily arXiv:1201.0002[astro-ph.IM]}}].

\bibitem{TelescopeArray:2014tsd}
{\scshape Telescope Array} collaboration, \emph{{Indications of Intermediate-Scale Anisotropy of Cosmic Rays with Energy Greater Than 57 EeV in the Northern Sky Measured with the Surface Detector of the Telescope Array Experiment}}, \href{https://doi.org/10.1088/2041-8205/790/2/L21}{\emph{Astrophys. J. Lett.} {\bfseries 790} (2014) L21} [\href{https://arxiv.org/abs/1404.5890}{{\ttfamily arXiv:1404.5890[astro-ph.HE]}}].

\bibitem{Li:1983fv}
T.P.~Li and Y.Q.~Ma, \emph{{Analysis methods for results in gamma-ray astronomy}}, \href{https://doi.org/10.1086/161295}{\emph{Astrophys. J.} {\bfseries 272} (1983) 317}.

\bibitem{He:2014mqa}
H.-N.~He, A.~Kusenko, S.~Nagataki, B.-B.~Zhang, R.-Z.~Yang and Y.-Z.~Fan, \emph{{Monte Carlo Bayesian search for the plausible source of the Telescope Array hotspot}}, \href{https://doi.org/10.1103/PhysRevD.93.043011}{\emph{Phys. Rev. D} {\bfseries 93} (2016) 043011} [\href{https://arxiv.org/abs/1411.5273}{{\ttfamily arXiv:1411.5273[astro-ph.HE]}}].

\bibitem{Fang:2014uja}
K.~Fang, T.~Fujii, T.~Linden and A.V.~Olinto, \emph{{Is the Ultra-High Energy Cosmic-Ray Excess Observed by the Telescope Array Correlated with IceCube Neutrinos?}}, \href{https://doi.org/10.1088/0004-637X/794/2/126}{\emph{Astrophys. J.} {\bfseries 794} (2014) 126} [\href{https://arxiv.org/abs/1404.6237}{{\ttfamily arXiv:1404.6237[astro-ph.HE]}}].

\bibitem{Kim:2019eib}
J.~Kim, D.~Ryu, H.~Kang, S.~Kim and S.-C.~Rey, \emph{{Filaments of galaxies as a clue to the origin of ultrahigh-energy cosmic rays}}, \href{https://doi.org/10.1126/sciadv.aau8227}{\emph{Sci. Adv.} {\bfseries 5} (2019) eaau8227} [\href{https://arxiv.org/abs/1901.00627}{{\ttfamily arXiv:1901.00627[astro-ph.HE]}}].

\bibitem{TelescopeArray:2021dfb}
{\scshape Telescope Array} collaboration, \emph{{Indications of a Cosmic Ray Source in the Perseus-Pisces Supercluster}},  \href{https://arxiv.org/abs/2110.14827}{{\ttfamily arXiv:2110.14827[astro-ph.HE]}}.

\bibitem{Kim:2023xfc}
{\scshape Telescope Array} collaboration, \emph{{Updates on the Hotspot and the Perseus-Pisces supercluster Excess Observed by the Telescope Array Experiment}}, \href{https://doi.org/10.1051/epjconf/202328303005}{\emph{EPJ Web Conf.} {\bfseries 283} (2023) 03005}.

\bibitem{Kim:2023ksw}
{\scshape Telescope Array} collaboration, \emph{{Anisotropies in the arrival direction distribution of ultra-high energy cosmic rays measured by the Telescope Array surface detector}}, \href{https://doi.org/10.22323/1.444.0244}{\emph{PoS} {\bfseries ICRC2023} (2023) 244}.

\bibitem{PierreAuger:2007pcg}
{\scshape Pierre Auger} collaboration, \emph{{Correlation of the highest energy cosmic rays with nearby extragalactic objects}}, \href{https://doi.org/10.1126/science.1151124}{\emph{Science} {\bfseries 318} (2007) 938} [\href{https://arxiv.org/abs/0711.2256}{{\ttfamily arXiv:0711.2256[astro-ph]}}].

\bibitem{AGASA:1998frs}
{\scshape AGASA} collaboration, \emph{{The Anisotropy of cosmic ray arrival directions around 10**18-eV}}, \href{https://doi.org/10.1016/S0927-6505(98)00064-4}{\emph{Astropart. Phys.} {\bfseries 10} (1999) 303} [\href{https://arxiv.org/abs/astro-ph/9807045}{{\ttfamily astro-ph/9807045}}].

\bibitem{AGASA:1999jxd}
{\scshape AGASA} collaboration, \emph{{The anisotropy of cosmic ray arrival direction around 10\textasciicircum{}18ev}},  in \emph{{26th International Cosmic Ray Conference}}, 6, 1999 [\href{https://arxiv.org/abs/astro-ph/9906056}{{\ttfamily astro-ph/9906056}}].

\bibitem{TelescopeArray:2018ygn}
{\scshape Telescope Array} collaboration, \emph{{Evidence for Declination Dependence of the Ultrahigh Energy Cosmic Ray Spectrum in the Northern Hemisphere}},  \href{https://arxiv.org/abs/1801.07820}{{\ttfamily arXiv:1801.07820[astro-ph.HE]}}.

\bibitem{TelescopeArray:2024tbi}
{\scshape Telescope Array} collaboration, \emph{{Observation of Declination Dependence in the Cosmic Ray Energy Spectrum}},  \href{https://arxiv.org/abs/2406.08612}{{\ttfamily arXiv:2406.08612[astro-ph.HE]}}.

\bibitem{Courtois:2013yfa}
H.M.~Courtois, D.~Pomarede, R.B.~Tully and D.~Courtois, \emph{{Cosmography of the Local Universe}}, \href{https://doi.org/10.1088/0004-6256/146/3/69}{\emph{Astron. J.} {\bfseries 146} (2013) 69} [\href{https://arxiv.org/abs/1306.0091}{{\ttfamily arXiv:1306.0091[astro-ph.CO]}}].

\bibitem{Tully:2019ngb}
R.B.~Tully, D.~Pomarede, R.~Graziani, H.M.~Courtois, Y.~Hoffman and E.J.~Shaya, \emph{{Cosmicflows-3: Cosmography of the Local Void}}, \href{https://doi.org/10.3847/1538-4357/ab2597}{\emph{Astrophys. J.} {\bfseries 880} (2019) 24} [\href{https://arxiv.org/abs/1905.08329}{{\ttfamily arXiv:1905.08329[astro-ph.CO]}}].

\end{thebibliography}\endgroup


{\small

}

%\input{main.bbl}

\newpage
\section*{Full Authors List: Telescope Array Collaboration}

\makeatletter
\newcommand{\ssymbol}[1]{^{\@fnsymbol{#1}}}
\makeatother
\par\noindent
R.U.~Abbasi$^{1}$,
T.~Abu-Zayyad$^{1,2}$,
M.~Allen$^{2}$,
J.W.~Belz$^{2}$,
D.R.~Bergman$^{2}$,
F.~Bradfield$^{3}$,
I.~Buckland$^{2}$,
W.~Campbell$^{2}$,
B.G.~Cheon$^{4}$,
K.~Endo$^{3}$
% ...
\bigskip
\par\noindent
{\footnotesize\it
$^{1}$ Department of Physics, Loyola University-Chicago, Chicago, Illinois 60660, USA \\
$^{2}$ High Energy Astrophysics Institute and Department of Physics and Astronomy, University of Utah, Salt Lake City, Utah 84112-0830, USA \\
$^{3}$ Graduate School of Science, Osaka Metropolitan University, Sugimoto, Sumiyoshi, Osaka 558-8585, Japan \\
$^{4}$ Department of Physics and The Research Institute of Natural Science, Hanyang University, Seongdong-gu, Seoul 426-791, Korea \\
% ...

\let\thefootnote\relax\footnote{$\ssymbol{1}$ Presently at: Purple Mountain Observatory, Nanjing 210023, China}
\let\thefootnote\relax\footnote{$\ssymbol{2}$ Presently at: Physics Department, Brookhaven National Laboratory, Upton, NY 11973, USA}
\let\thefootnote\relax\footnote{$\ssymbol{3}$ Presently at: Korea Institute of Geoscience and Mineral Resources, Daejeon, 34132, Korea}
\let\thefootnote\relax\footnote{$\ssymbol{4}$ Presently at: NASA Marshall Space Flight Center, Huntsville, Alabama 35812, USA}
\addtocounter{footnote}{-1}\let\thefootnote\svthefootnote
}
\par\noindent

\section*{Acknowledgements}

The Telescope Array experiment is supported by the Japan Society for
the Promotion of Science(JSPS) through
Grants-in-Aid
for Priority Area
%"Highest Energy Cosmic Rays"
431,
for Specially Promoted Research
%``Extreme Phenomena in the Universe Explored by Highest Energy Cosmic Rays''
%Grant Number
JP21000002,
%Grant-in-Aid
for Scientific  Research (S)
%"Quest for the unified picture of the explosion mechanism of supernovae and the central engine of gamma-ray bursts"
%Grant Number
JP19104006,
%Grant-in-Aid
for Specially Promoted Research
%"Extended Telescope Array Experiment - Nearby Extreme Universe Elucidated by Highest-energy Cosmic Rays"
%Grant Number
JP15H05693,
%Grant-in-Aid
for Scientific  Research (S)
%Grant Number
JP19H05607,
%Grant-in-Aid
for Scientific  Research (S)
%"Study of the ultra high energy cosmic ray source evolution by detailed measurement of cosmic rays in the wide energy range"
%Grant Number
JP15H05741,
%Grant-in-Aid
for Science Research (A)
%Grant Number
JP18H03705,
%Grant-in-Aid
for Young Scientists (A)
%"hoge hoge"
%Grant Number
JPH26707011,
%Grant-in-Aid
for Transformative Research Areas (A)
% Transformative Research Areas, Section (II)
%"Probing Space and Earth Environments via Cosmic-Ray Muon Imaging"
%Grant Number
JP25H01294,
%Grant-in-Aid
for International Collaborative Research
%Grant Number
24KK0064,
and for Fostering Joint International Research (B)
%"Search for Ultra-High Energy Cosmic Ray origin using the extended Telescope Array experiment"
%Grant Number
JP19KK0074,
by the joint research program of the Institute for Cosmic Ray Research (ICRR), The University of Tokyo;
by the Pioneering Program of RIKEN for the Evolution of Matter in the Universe (r-EMU);
by the U.S. National Science Foundation awards
PHY-1806797, PHY-2012934, PHY-2112904, PHY-2209583, PHY-2209584, and PHY-2310163, as well as AGS-1613260, AGS-1844306, and AGS-2112709;
by the National Research Foundation of Korea
% \linebreak
(2017K1A4A3015188, 2020R1A2C1008230, and RS-2025-00556637) ;
%\linebreak
by the Ministry of Science and Higher Education of the Russian Federation under the contract 075-15-2024-541, IISN project No. 4.4501.18, by the Belgian Science Policy under IUAP VII/37 (ULB), by National Science Centre in Poland grant 2020/37/B/ST9/01821, by the European Union and Czech Ministry of Education, Youth and Sports through the FORTE project No. CZ.02.01.01/00/22\_008/0004632, and by the Simons Foundation (MP-SCMPS-00001470, NG). This work was partially supported by the grants of the joint research program of the Institute for Space-Earth Environmental Research, Nagoya University and Inter-University Research Program of the Institute for Cosmic Ray Research of University of Tokyo. The foundations of Dr. Ezekiel R. and Edna Wattis Dumke, Willard L. Eccles, and George S. and Dolores Dor\'e Eccles all helped with generous donations. The State of Utah supported the project through its Economic Development Board, and the University of Utah through the Office of the Vice President for Research. The experimental site became available through the cooperation of the Utah School and Institutional Trust Lands Administration (SITLA), U.S. Bureau of Land Management (BLM), and the U.S. Air Force. We appreciate the assistance of the State of Utah and Fillmore offices of the BLM in crafting the Plan of Development for the site.  We thank Patrick A.~Shea who assisted the collaboration with much valuable advice and provided support for the collaboration’s efforts. The people and the officials of Millard County, Utah have been a source of steadfast and warm support for our work which we greatly appreciate. We are indebted to the Millard County Road Department for their efforts to maintain and clear the roads which get us to our sites. We gratefully acknowledge the contribution from the technical staffs of our home institutions. An allocation of computing resources from the Center for High Performance Computing at the University of Utah as well as the Academia Sinica Grid Computing Center (ASGC) is gratefully acknowledged.

\end{document}